# A critical analysis of the modelling of dissipation in fission[a]

B. Jurado[1][b], C. Schmitt[1], K.-H. Schmidt[1], J. Benlliure[2], A. R. Junghans[3]

[1)] *GSI, Planckstr.1, 64291 Darmstadt, Germany*
[2)] *Univ. de Santiago de Compostela, 15706 S. de Compostela, Spain*
[3)] *Forschungszentrum Rossendorf, Postfach 510119, 01314 Dresden, Germany*

**Abstract:** The time-dependent flux over the fission barrier of an excited nucleus under the influence of dissipation is investigated. Characteristic features of the evolution of the amplitude of the probability distribution and the velocity profile at the fission barrier are derived. Analytical results are compared to numerical Langevin calculations and used to develop a new analytical approximation to the solution of the Fokker-Planck equation for the time-dependent fission-decay width. This approximation is shown to be more realistic than previously proposed descriptions, which were widely used in the past..

## 1. Introduction

One of the topical features of nuclear dynamics during the last decades is the role of dissipation for such processes as deep-inelastic heavy-ion collisions [1], damping of giant resonances [2] and induced fission [3] – to learn whether collective motion in nuclei is under-damped like in water or over-damped like in honey droplets. In spite of intensive experimental and theoretical work, most conclusions on the dissipation strength are not well established. The situation still remains unclear concerning the deformation and temperature dependence of nuclear friction as well. From the theoretical point of view, different models have been developed, e.g. the one-body dissipation [4] concept based on the wall-and-window formula, the two-body viscosity model [5] or quantum transport theories [6, 7, 8]. They yield different results for the magnitude as well as for the dependence on temperature and deformation of the dissipation strength.

Although the role of nuclear dissipation in fission has been recognized long time ago, fission continues to be one of the most promising tools for deducing quantitative conclusions. More than 60 years ago, Bohr and Wheeler [9] proposed a derivation of the fission decay width $\Gamma_f^{BW}$ based on purely statistical considerations in the framework of the transition-state model. Soon after, in 1940, Kramers [10] developed the first transport theory to describe nuclear fission. In his model, fission is described as the evolution of the collective degrees of freedom in interaction with a heat bath formed by the single nucleons. This brings in the concept of dissipation, which represents the transformation of collective motion into heat due to the damping mechanism. The description of Kramers was derived from the *stationary* solution of the Fokker-Planck equation (FPE). Therefore, it only holds after the system has reached

---
[a)] This work forms part of the PhD thesis of B. Jurado
[b)] Corresponding author: B. Jurado, GANIL, Boulevard Henry Becquerel, B.P. 5027, 14076 Caen CEDEX 5, France, e-mail: jurado@ganil.fr (present address)

equilibrium. The success of the transition-state model of Bohr and Wheeler prevented this idea of Kramers to establish. Approximately forty years later, experimentally observed high pre-scission neutron multiplicities [11, 12] gave the impetus to Grangé et al. [13] to theoretically investigate the influence of dissipation on the fission time scale. Their numerical solution of the *time-dependent* FPE shows that it takes some time, the so-called *transient time* $\tau_{trans}$, until the current over the saddle point reaches its stationary value. Grangé et al. [13] pointed out that this transient time $\tau_{trans}$ but also an additional saddle-to-scission time $\tau_{ss}$ lead to an increase of pre-scission particle multiplicities. The transient time originates from the time needed by the probability distribution of the particle $W(x,v;t)$ (deformation $x$ and conjugate momentum $p = v \cdot \mu$, where $v$ is the velocity of the system and $\mu$ its inertia) to spread out in deformation space. From their numerical calculations, the authors of ref. [13] extracted the following approximation for the transient time, defined as the time until the fission width $\Gamma_f(t)$ reaches 90% of its asymptotic value:

$$\tau_{trans} = \frac{1}{\beta} \ln\left(\frac{10 B_f}{T}\right) \quad \text{for } \beta < 2\omega_g$$

$$\tau_{trans} = \frac{\beta}{2\omega_g^2} \ln\left(\frac{10 B_f}{T}\right) \quad \text{for } \beta > 2\omega_g$$

(1)

where $B_f$ is the fission-barrier height, $T$ is the nuclear temperature, $\omega_g$ is the effective oscillator frequency at the ground state, and $\beta$ is the reduced dissipation coefficient which measures the relative rate with which the excitation energy is transferred between the collective and intrinsic degrees of freedom.

Later, full dynamical calculations were performed with a stochastic approach based on the multidimensional time-dependent FPE [14] or Langevin equation [15, 16, 17, 18] with allowance for evaporation during the dynamical evolution of the system. These are two equivalent methods, corresponding to the integral, respectively differential, formulation of the same process. Whereas the solution of the FPE leads to the probability distribution $W(x,v;t)$ of the particle as a function of time, the Langevin approach consists of following the trajectory of every individual nucleus all along its path to fission. Because the Fokker-Planck approach gives directly access to the probability distribution $W(x,v;t)$, and consequently to the evolution of the fission decay width $\Gamma_f(t)$ with time (see equation 9), it corresponds to a more transparent way to get information on transient effects. However, for realistic physical cases the FPE can only be solved numerically. The same information can be extracted from a Langevin treatment as well, but only after some average over a large amount of trajectories. The possibility of the Langevin method to follow individual trajectories may explain why it is often preferred to the Fokker-Planck approach since several years.

Another procedure widely used to study fission dynamics consists of introducing a time-dependent fission decay width $\Gamma_f(t)$ in an evaporation code [19, 20, 21, 22, 23, 24]. In section 3 we will point out that such a treatment is *exactly equivalent* to solving the above-mentioned equation of motion with allowance for evaporation *under the condition* that the used time-dependent width $\Gamma_f(t)$ is obtained by the Fokker-Planck or Langevin solution at each evaporation step. Thus, it takes into account the changes in mass, charge, excitation energy and angular momentum of the decaying system. Unfortunately, following a large amount of trajectories or solving numerically the FPE at each evaporation step needs a very high computational effort, which may exceed the technical possibilities actually available for many applications. However, in an evaporation code one could get around the high computing time needed to calculate numerically the time-dependent fission width at each evaporation step by using a suitable analytically calculable expression for $\Gamma_f(t)$.

With this aim in view, the main task of the present work is to shed light on the characteristic features of transient effects in the dynamical evolution of the nuclear system. On this basis, we propose a way to model the influence of dissipation in nuclear fission in terms of a realistic analytical approximation for $\Gamma_f(t)$. In the following paper [25], we investigate, how transient effects manifest. There we will make use of peripheral nuclear collisions at relativistic energies as an appropriate reaction mechanism dedicated to dissipation studies and establish the requirements on relevant experimental observables that are sensitive to transient effects.

This work is also motivated by the need of incorporating realistic features of fission dynamics in complex model calculations for technical applications, e.g. the nuclide production in secondary-beam facilities, in spallation-neutron sources, in the core of an accelerator-driven system, and in shielding calculations. In these codes, the computational effort is already very high due to the necessary transport calculations in a thick-target environment, and thus the explicit solution of the equation of motion, e.g. by the use of the Langevin approach, seems to be excluded.

## 2. Time evolution of the fission-decay width under the influence of dissipation

### 2.1. Previously used approximations of the fission-decay width

If the initial population of the nuclear system in deformation and conjugate momentum differs from equilibrium, the system is subject to relaxation effects. The influence of these effects on the time dependence of the fission-decay width has carefully been studied by Grangé, Weidenmüller and collaborators [13, 26] on the basis of the numerical solution of the FPE. As an example, we show their result for the evolution of the escape rate $\lambda_f(t) = \Gamma_f(t)/\hbar$ of the compound nucleus $^{248}$Cm at a temperature of T = 5 MeV with a friction coefficient of $\beta = 5 \cdot 10^{21}$ s$^{-1}$ in Figure 1. The potential used is specified in Figure 2. The fission-decay width as a function of time can be characterised by three main features: a delayed onset, a rising part and a stationary value. We will see that the initial suppression of the fission width has a decisive influence on the evolution of the system. Indeed, the inhibition of fission during the transient time $\tau_{trans}$ increases the chance of the nucleus to decay at the earliest times by particle emission. The resulting loss of excitation energy and the change of the properties (e.g. fissility) of the residual nucleus reduce its fission probability.

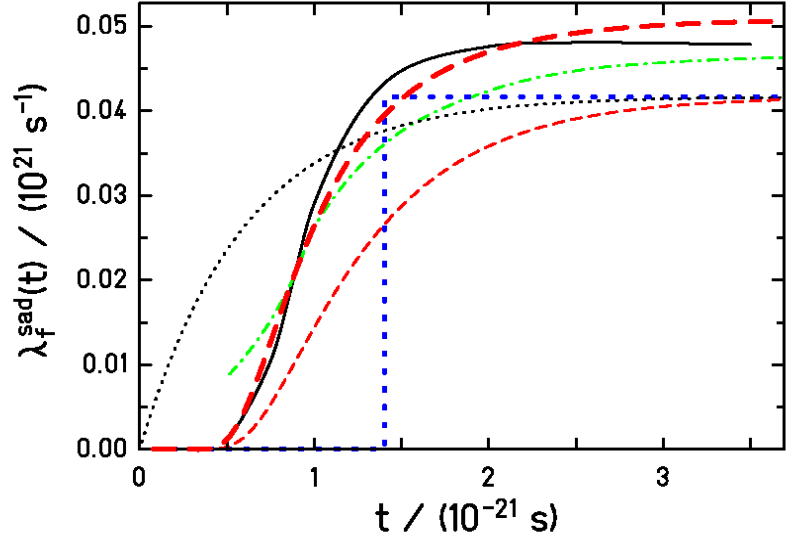

**Figure 1**: The solution of the Fokker-Planck equation (full line) for the time-dependent escape rate for the schematic case of a fissioning $^{248}$Cm nucleus, using the potential introduced by Bhatt et al. [26] (see Figure 2) at a temperature $T = 5$ MeV and for a reduced dissipation coefficient $\beta = 5 \cdot 10^{21}$ s$^{-1}$. The initial condition corresponds to equilibrium at a temperature $T_{initial} = 0.3$ MeV, introduced in ref. [26] to represent the quantum-mechanical zero-point motion in the ground state. This solution is compared with different approximations (the step function [27]: thick dotted line, the exponential-like function [28]: thin dotted line, the approximate formulation of Bhatt et al. [26]: dash-dotted line, the approximate formulation of ref. [30]: thin dashed line, and the improved expression proposed in this work: thick dashed line). The solution of the FPE and the approximation of Bhatt et al. are taken from Figure 2 of ref. [26], where values for $t < 0.5 \cdot 10^{-21}$ s are not shown.

In the past, several approximations for the time evolution of the fission-decay width have been proposed. The two most widely used are:

- a step function [27] that sets in at the transient time $\tau_{trans}$:

$$\Gamma_f(t) = \begin{cases} 0, & t < \tau_{trans} \\ \Gamma_f^k, & t \geq \tau_{trans} \end{cases} \quad (2)$$

- an exponential in-growth function [28] defined by:

$$\Gamma_f(t) = \Gamma_f^k \cdot \{1 - \exp(-t/\tau)\} \quad (3)$$

where : $\tau = \tau_{trans}/2.3$ and $\Gamma_f^K$ is the Kramers decay width:

$$\Gamma_f^K = K \cdot \Gamma_f^{BW} \quad (4)$$

The fission-decay width of the transition-state model $\Gamma_f^{BW}$ may conveniently be expressed by the following approximation, introduced by Moretto [29]:

$$\Gamma_f^{BW} \approx \frac{1}{2\pi\rho(E)} T_{sad} \rho_{sad}(E - B_f) \quad (5)$$

and the Kramers factor $K$ is given by

$$K = \left[1 + \left(\frac{\beta}{2\omega_{sad}}\right)^2\right]^{1/2} - \frac{\beta}{2\omega_{sad}} \quad (6)$$

where: $\omega_{sad}$ is the frequency of the harmonic-oscillator potential that osculates the fission barrier at the saddle point, $\rho_{sad}(E)$ and $T_{sad}$ are the level density and temperature of the fissioning nucleus at saddle, respectively, and $\rho(E)$ is the level density in the ground state.

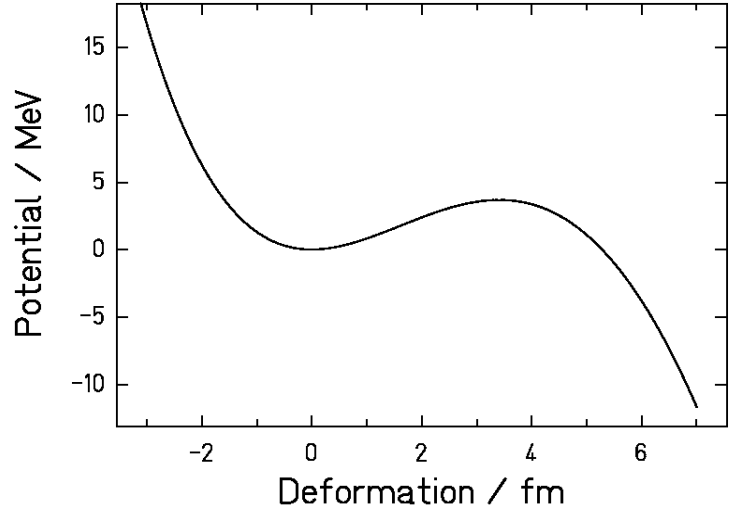

**Figure 2**: Presentation of the potential in fission direction of $^{248}$Cm as given by Bhatt et al. [26]: ($V = 8.61\cdot10^{-3}(x-3.41)^2\cdot[(x-23.098)(x+1.59)]+3.7$). The fission-barrier height is 3.7 MeV.

A more elaborate formulation of $\Gamma_f(t)$ was derived by Grangé et al. [13] and improved by Bhatt et al. [26] a few years later. Using the Gaussian approximation, they derived an analytical solution of the FPE in the under-damped regime and an analytical solution of the Smoluchowski equation in the over-damped case. Recently, we have proposed another expression for $\Gamma_f(t)$ in ref. [30]. In the present work, we will present a refined version of this description (see section 2.5). All these approximations are compared to the exact solution of the FPE in Figure 1. The exponential-like in-growth function shows strongly rising values already at very early time, which is in contrast to the numerical solution of the FPE. Even if the step function is able to describe the inhibition of fission for small times, this suppression is certainly too strong, and its steep slope is too crude. The two rather elaborate approximations of ref. [26] and [30] are better suited. However, while the formulation of Bhatt et al. [26] overestimates the fission-decay rate at early times, the onset of $\Gamma_f(t)$ is well described by our approximation [30] and its improved expression proposed in this work. The curves also slightly differ in their asymptotic values. In view of these important differences between the various approximations, it is legitimate to investigate what are the most relevant characteristics for a realistic description of dissipative effects in fission.

## 2.2. Features of the relaxation process

In this section, some basic features of the relaxation process towards equilibrium are carefully investigated with the help of Langevin calculations. We take the case of $^{248}$Cm introduced by Bhatt et al. [26] as an example. The discretisation method we use in this work to solve the Langevin equation is documented in appendix **A2**.

The dynamics of a fissioning system is well characterised by the time-dependent flux of trajectories in deformation space. The time-dependent fission decay width $\Gamma_f(t)$ can easily be derived from the flux $\phi_{x_b}(t)$ evaluated at the saddle point $x_b$, which represents the current across the fission barrier:

$$\phi_{x_b}(t) = \int_{-\infty}^{+\infty} v \cdot W(x_b, v; t) dv \tag{7}$$

Let

$$\Pi(x_b; t) = \int_{-\infty}^{+x_b} dx \int_{-\infty}^{+\infty} W(x, v; t) dv \tag{8}$$

be the probability that the system is at deformations $x < x_b$.

The time-dependent fission width $\Gamma_f(t)$, related to the escape rate $\lambda_f(t)$, is then defined by:

$$\Gamma_f(t) = \hbar \lambda_f(t) = \hbar \frac{\phi_{x_b}(t)}{\Pi(x_b; t)} = \hbar \frac{\int_{-\infty}^{+\infty} v \cdot W(x_b, v; t) dv}{\int_{-\infty}^{+x_b} dx \int_{-\infty}^{+\infty} W(x, v; t) dv} \tag{9}$$

By introducing the mean velocity at the barrier $\bar{v}(x = x_b; t)$ into the definition of $\phi_{x_b}(t)$, we see that the flux at the barrier can be expressed as the product of the mean velocity $\bar{v}(x = x_b;t)$ and the amplitude of the probability distribution $\int_{-\infty}^{+\infty} W(x = x_b,v;t)dv$ at the barrier $x_b$:

$$\phi_{x_b}(t) = \frac{\int_{-\infty}^{+\infty} v \cdot W(x = x_b,v;t)dv}{\int_{-\infty}^{+\infty} W(x = x_b,v;t)dv} \cdot \int_{-\infty}^{+\infty} W(x = x_b,v;t)dv = \bar{v}(x = x_b;t) \cdot \int_{-\infty}^{+\infty} W(x = x_b,v;t)dv \qquad (10)$$

Let us investigate the time evolution of the two terms of equation (10). Figure 3 represents the velocity distribution at the saddle point $W(x = x_b,v;t)$ as a function of time, and Figure 4 shows the variation with time of the amplitude of the probability distribution $\int_{-\infty}^{+\infty} W(x = x_b,v;t)dv$ at the barrier. Both quantities were obtained by a Langevin calculation. As can be seen in Figure 3, the mean velocity at the barrier $\bar{v}(x = x_b;t)$ gradually decreases during the onset of the fission width before reaching its asymptotic value in the stationary state. However, the variation of the amplitude of the probability distribution $\int_{-\infty}^{+\infty} W(x = x_b,v;t)dv$ is much more important. Indeed the variation of the amplitude with time extends over many orders of magnitude during the transient time as exhibited in Figure 4. Therefore, the variation of the amplitude of the probability distribution at the fission barrier represents the dominating influence on the onset of the flux, and consequently of the decay width.

For a more general understanding of the respective importance of the different contributions, we would like to compare the influences of the amplitude and of the mean velocity on the flux in the simple example of diffusion (over-damped motion) without driving force. In this limit, the FPE leads to the following reduced Smoluchowski equation [31]:

$$\frac{\partial}{\partial t} W(x;t) = +\frac{\partial^2}{\partial x^2}\left(\frac{T}{\beta\mu}\right) W(x;t) \qquad (11)$$

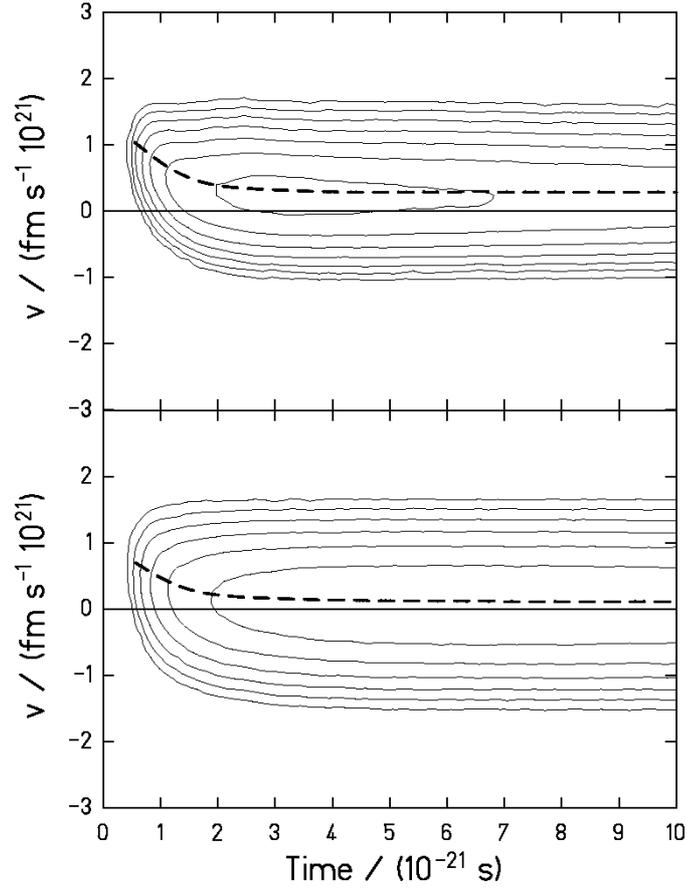

**Figure 3:** Two-dimensional contour plot of the velocity distribution at the barrier $W(x = x_b, v; t)$ as a function of time. Contour lines mark heights separated by a factor of two. The result has been obtained from a numerical calculation using the Langevin approach with the parameters $T = 3$ MeV and $\beta = 1 \cdot 10^{21}$ s$^{-1}$ (upper figure) and $T = 3$ MeV and $\beta = 5 \cdot 10^{21}$ s$^{-1}$ (lower figure). The calculation starts with the distribution in deformation $x$ and conjugate momentum $p = \mu v$ given by the zero-point motion in the ground state. The dashed line represents the mean velocity at the barrier as a function of time.

The probability distribution with initial width zero in the ground state at $x = 0$ and initial velocity $v = 0$ thus evolves in the following way [32]:

$$W(x;t) = \frac{1}{\sqrt{2\pi}\sigma_x} \cdot \exp\left(-x^2/\left(2\sigma_x^2\right)\right) \quad (12)$$

with

$$\sigma_x^2 = \frac{2T}{\mu\beta} \cdot t \quad (13)$$

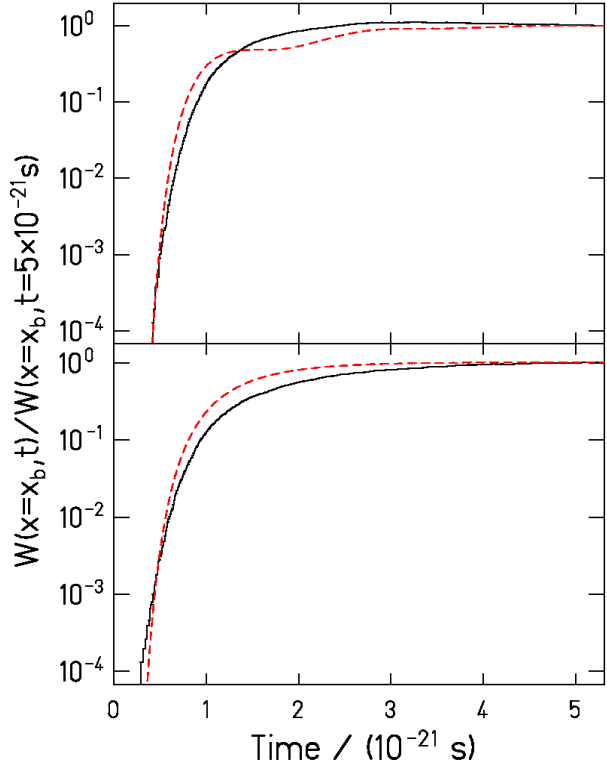

**Figure 4**: Amplitude of the probability distribution at the barrier deformation $x_b$ from the Langevin calculation (full line) above the potential of Figure 2 with the parameters $T = 3$ MeV and $\beta = 1 \cdot 10^{21}$ s$^{-1}$ (upper part), respectively $\beta = 5 \cdot 10^{21}$ s$^{-1}$ (lower part), compared with the amplitude at the barrier deformation $x_b$ above the parabola adapted to the curvature of the potential in the ground state (dashed line). All functions are normalised to the value at $t = 5 \cdot 10^{-21}$ s.

The amplitude W($x_b$;t) of the system at the barrier is thus given by:

$$W(x_b;t) = \frac{1}{\sqrt{2\pi}\sigma_x} \cdot \exp(-x_b^2/(2\sigma_x^2)) = \sqrt{\frac{\beta\mu}{4\pi T}} \frac{1}{\sqrt{t}} \exp\left(-\frac{x_b^2 \beta\mu}{4Tt}\right) \propto \frac{1}{\sqrt{t}} \exp\left(-\frac{1}{t}\right) \quad (14)$$

By applying the continuity equation (see also equation (36)), one obtains the mean velocity $\bar{v}$ at the barrier:

$$\bar{v}(x_b;t) = \frac{T}{\beta\mu} \frac{x_b}{\sigma_x^2} = \frac{x_b}{2t} \propto \frac{1}{t} \quad (15)$$

Thus, the flux as the product of amplitude and velocity varies like $\frac{1}{t^{3/2}} \exp\left(-\frac{1}{t}\right)$. Since for small times $\frac{1}{t^n} \exp\left(-\frac{1}{t}\right)$ behaves like $\exp\left(-\frac{1}{t}\right)$, $\forall n$, the amplitude governs the evolution of the flux at the beginning of the process as illustrated in Figure 5.

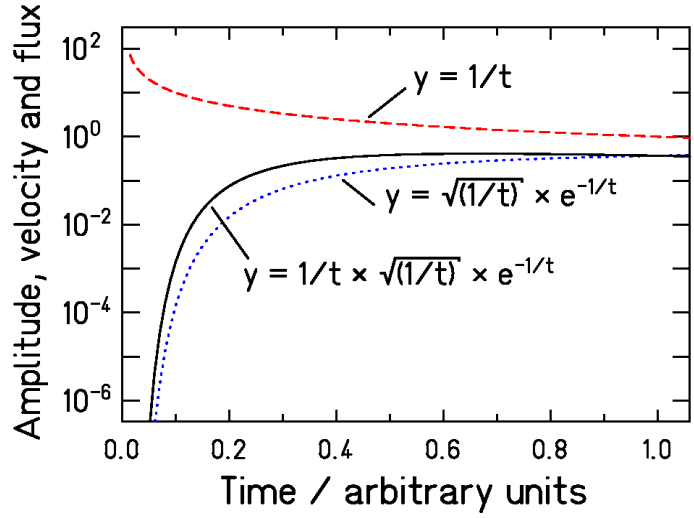

**Figure 5:** Schematic comparison of the influence of amplitude and velocity on the flux at the barrier in a simple diffusion problem. See text for details.

Let us now consider the influence of the shape of the potential on the time dependence of the amplitude at the barrier. With this aim, Figure 4 compares the variation of the amplitude at the barrier $W(x = x_b; t)$ obtained *numerically* using the realistic potential of Figure 2 with the solution of the FPE obtained *analytically* for a parabolic potential. The curvature of the latter corresponds to the curvature of the realistic potential in the ground state. One observes that both curves vary over many orders of magnitude in a very similar way. This result, perhaps surprisingly, suggests the minor impact of the shape of the potential on the variation of the amplitude.

In addition to these numerical results, we would like to strengthen this conclusion by more analytical arguments. With this aim in view, we compare the result for a simple diffusion problem without driving force obtained by solving the Smoluchowski equation (11) and the result for the parabolic potential in the over-damped region at a given deformation $\Delta x$. In both cases, the distribution is a Gaussian function in deformation given above by equation (12). The solution for $\sigma_x$ as a function of time for the first diffusion case was already given in equation (13)

$$\sigma_x^2 = \frac{2T}{\mu\beta} \cdot t$$

while the full solution for a parabolic potential [32] corresponds to

$$\sigma_x^2 = \frac{T}{\mu\omega_g^2}\left\{1 - \exp(-\beta t)\left[\frac{2\beta^2}{\beta_1^2}\sinh^2\left(\frac{\beta_1 t}{2}\right) + \frac{\beta}{\beta_1}\sinh(\beta_1 t) + 1\right]\right\} \quad (16)$$

where: $$\beta_1 = (\beta^2 - 4\omega_g^2)^{1/2} \quad (17)$$

In the over-damped regime, the physical meaning of equation (16) may be easily understood: $\beta$ is large and equation (16) can be approximated by the following equation

$$\sigma_x^2 \approx \frac{T}{\mu \omega_g^2} \left\{ 1 - \exp\left( -\frac{2\omega_g^2}{\beta} \cdot t \right) \right\} \qquad (18)$$

in which we have replaced $\beta_1 = \beta - 2\omega_g^2/\beta$ by $\beta$ and neglected the terms $\exp(-\beta t)$. Actually, equation (16) is already quantitatively very similar to equation (18) for $\beta \geq 5 \cdot 10^{21}$ s$^{-1}$. Equation (18) shows that the process of the population of the deformation space can be described by a probability distribution with the shape of a Gaussian whose second moment exponentially approaches the asymptotic value.

We see that equation (13) is the first-order approximation of equation (18), indicating that up to $t \approx \frac{\beta}{2\omega_g^2}$ the evolution of the amplitude of the probability distribution at the barrier deformation for a parabolic potential is very similar to that for the simple diffusion problem. Since the transient time $\tau_{trans}$ in the over-damped regime as defined by Grangé et al. [13] is approximately given by equation (1), the above-mentioned similarity extends to around half the transient time. We conclude that the spreading of the probability distribution during the time lapse in which the flux sets in is rather insensitive to the shape of the potential.

In Figure 4, we have compared the amplitude obtained for a realistic and a parabolic potential because the FPE can be solved analytically only in the simplest cases for which the potential has a parabolic form and the transport coefficients are constant. More complicated situations, e.g. for a realistic potential, like the one displayed in Figure 2, or in the case of deformation-dependent transport coefficients, have to be solved numerically. Concerning the approximation of transport coefficients that do not depend on deformation, the importance of such an assumption can be considerably reduced by the choice of an appropriate coordinate system as demonstrated in appendix **A1**. There it is shown that any given problem can be reformulated in an adapted coordinate system, where the inertia or friction coefficient does not vary with deformation, without changing the physics of the problem. In addition it is pointed out that theoretical one-body and two-body dissipation models predict that the variation of the dissipation strength with deformation is weak.

The initial conditions of the system concern the population in deformation and conjugate momentum as well as the temperature and angular momentum of the nucleus. Their determination relies on the description of the initial properties of the nuclear system after the reaction to be studied. Therefore, the initial population depends strongly on the entrance channel considered. While e.g. fusion reactions can lead to rather deformed compound nuclei and involve a broad range of angular momentum, very peripheral nuclear collisions at relativistic energies are characterised by small initial shape distortions and low angular momenta. The initial conditions are important for the evolution of the system, because they affect the time dependence of the fission-decay width $\Gamma_f(t)$. In section **2.4** we will propose a way to take the initial conditions into account in the analytical approximation of the time-dependent fission width $\Gamma_f(t)$.

## 2.3. New description of the time-dependent fission-decay width $\Gamma_f(t)$

In the previous section, we have collected some characteristics of the dissipation process, which have a rather general validity: the predominance of the amplitude for the time evolution of the flux and the minor importance of the shape of the potential for the early evolution of the probability distribution in deformation and conjugate momentum. In ref. [30] a realistic approximation for the decay width $\Gamma_f(t)$ that is based on these features has been presented. In the following, the various steps to derive this approximation will be illustrated in detail.

First of all, we will develop a slightly modified expression for the exact description of the fission-decay width, whose definition was already given above in equation (9). We start by defining the normalised probability distribution $W_n(x=x_b,v,t)$ at the fission-barrier deformation $x_b$ as:

$$W_n(x=x_b,v,t) = \frac{W(x=x_b,v,t)}{\int_{-\infty}^{x_b}\int_{-\infty}^{+\infty} W(x,v,t)dvdx} \qquad (19)$$

Considering equation (19) and introducing the mean velocity $\bar{v}$ at the barrier deformation already used in equation (10), the fission width of equation (9), can be reformulated as follows:

$$\Gamma_f(t) = \hbar \cdot \bar{v}(x=x_b;t) \cdot \int_{-\infty}^{+\infty} W_n(x=x_b,v,t)dv \qquad (20)$$

By defining the amplitude at the barrier, integrated over velocity by

$$W_n(x=x_b;t) = \int_{-\infty}^{+\infty} W_n(x=x_b,v;t)dv \qquad (21)$$

the fission width can be written as

$$\Gamma_f(t) = \hbar \cdot \bar{v}(x=x_b;t) \cdot W_n(x=x_b;t) \qquad (22)$$

In the stationary case we get:

$$\Gamma_f(t \to \infty) = \hbar \cdot \bar{v}(x=x_b;t \to \infty) \cdot W_n(x=x_b,t \to \infty) \approx \Gamma_f^K \qquad (23)$$

where we have identified the asymptotic fission width by the Kramers expression $\Gamma_f^K$ from equation (4).

Finally, combining equations (22) and (23), we can reformulate the time-dependent fission width as:

$$\Gamma_f(t) \approx \frac{\bar{v}(x=x_b;t)\cdot W_n(x=x_b;t)}{\bar{v}(x=x_b;t\to\infty)\cdot W_n(x=x_b;t\to\infty)}\cdot \Gamma_f^K \qquad (24)$$

At this point, we introduce two approximations that lead to a new analytical description of the time-dependent fission width. The first approximation is to neglect the variation of the mean velocity with time. Thus, $\bar{v}(x=x_b;t)$ is replaced by its asymptotic value $\bar{v}(x=x_b;t\to\infty)$ in the numerator of equation (24), leading to the following expression:

$$\Gamma_f(t) \approx \frac{W_n(x=x_b;t)}{W_n(x=x_b;t\to\infty)}\Gamma_f^K \qquad (25)$$

This approximation is well justified by our previous investigations in section **2.2** that have shown that the variation of the mean velocity at the barrier is small compared to the variation of the amplitude, and therefore the evolution of the amplitude $\int_{-\infty}^{+\infty} W(x=x_b,v;t)dv$ with time has a much stronger influence than the mean velocity.

A second approximation which we used in ref. [30] consists of expressing the shape of the in-growth function at the fission barrier, which is given by the shape of $W_n(x = x_b, t)$ in equation (25), by the analytical solution derived for a parabolic potential [32]. The validity of this second simplification is again justified by our previous investigations in section **2.2**, where we have shown that the amplitudes at the barrier $W$ and $W^{par}$ for a realistic and a parabolic potential, respectively, evolve in a quite similar way at the beginning of the process, see Figure 4. Consequently, the following set of equations represents an *analytical* expression of the fission decay width that is based on realistic assumptions:

$$\Gamma_f(t) \approx \frac{W^{par}(x=x_b,t)}{W^{par}(x=x_b,t\to\infty)}\Gamma_f^K \qquad (26)$$

in which we implement the parabolic solution (equations (12) and (16)):

$$W^{par}(x,v;t) = \frac{1}{\sqrt{2\pi}\sigma_x}\cdot \exp\left(-x^2/(2\sigma_x^2)\right)$$

$$\sigma_x^2 = \frac{T}{\mu\omega_g^2}\left\{1-\exp(-\beta t)\left[\frac{2\beta^2}{\beta_1^2}\sinh^2\left(\frac{\beta_1 t}{2}\right)+\frac{\beta}{\beta_1}\sinh(\beta_1 t)+1\right]\right\}$$

Note, that we replaced the normalised probability $W_n^{par}$ by the unnormalised quantity $W^{par}$ in equation (26), because in the case of the parabolic potential the probability distribution is confined, and thus $\int_{-x_b}^{+\infty} W^{par}(x;t) \approx \int_{-\infty}^{+\infty} W^{par}(x;t)$.

## 2.4. Initial conditions

The analytical solution of the FPE for a parabolic potential derived in [32], which we use in equation (16), refers to specific initial conditions corresponding to a $\delta$ function in deformation and momentum. This means that the initial deformation is in the minimum of the parabola, corresponding to the nuclear ground state, and the initial momentum is zero. However, as discussed in section **2.2.**, a realistic description should include the initial population related to the reaction under study. Due to the uncertainty principle, the initial probability distribution $W(x,v;t=0)$ differs from a $\delta$ function whatever the entrance channel is. To account for this effect, we introduced in ref. [30] a time shift $t_0$ in equation (16) for the standard deviation $\sigma_x^2$:

$$\sigma_x^2 = \frac{T}{\mu\omega_g^2}\left\{1-\exp(-\beta(t-t_0))\left[\frac{2\beta^2}{\beta_1^2}\sinh^2\left(\frac{\beta_1(t-t_0)}{2}\right)+\frac{\beta}{\beta_1}\sinh(\beta_1(t-t_0))+1\right]\right\} \quad (27)$$

The time shift $t_0$ accounts for the time needed for the probability distributions, expressed by equations (12) and (16), to establish the initial distribution in deformation space. One should stress that such a procedure is not restricted to any specific initial condition. In fact, it can be applied to any reaction, under the condition that the phase space populated by the entrance channel can be assumed to be Gaussian. An alternative way to consider a finite width in the initial conditions of the analytical solution of the FPE for the parabolic potential (also restricted to Gaussian-like distributions) has been developed recently by Boilley et al. [33].

In [30] we assumed an initial distribution in deformation and momentum, which corresponds to the zero-point motion at the ground state of the nucleus. This is the narrowest distribution compatible with the uncertainty principle. This condition, related to a nucleus with a distribution in deformation corresponding to the nuclear ground state[c], is approximately valid for reactions which introduce small shape distortions like peripheral collisions at relativistic energies [34]. These reactions will be investigated in the framework of dissipation studies in our following paper [25]. The time shift $t_0$ corresponding to the zero-point motion was given in ref. [30] by:

$$t_0 = \text{MAX}\left\{\frac{1}{\beta}ln\left(\frac{2T}{2T-\hbar\omega_g}\right),\frac{\hbar\beta}{4\omega_g T}\right\} \quad (28)$$

Equation (28) accounts for both the under-damped and the over-damped cases. In the under-damped case, the deformation and the momentum coordinate saturate at about the same time.

---

[c]) A recent publication [Phys. Lett. B 567 (2003) 189.] criticises the application of an effective time shift, which was proposed in our previous publication [30] for introducing the initial condition of the zero-point motion. The criticism is based on a misinterpretation of "$t_0$ as a kind of relaxation time to the equilibrium of the oscillator, as represented by the ground state". We stress here that this time shift $t_0$ does not account for the relaxation towards equilibrium but represents a mean to model the distribution corresponding to the zero-point motion quantum state.

Therefore, the time shift needed for the probability distribution, expressed by equations (12) and (16), to reach the width of the zero-point motion in deformation is equal to the time that the average energy of the collective degree of freedom needs to reach the value $E_0 = \frac{1}{2}\hbar\omega_g$ associated to the zero-point motion. Considering the energy transfer $E_t$ from the heat bath of the intrinsic excitations at temperature $T$ to the oscillation in deformation as given by the FPE, the effective time delay $t_0^{under}$ is determined by:

$$E_t = T \cdot \left[1 - \exp\left(-\beta \cdot t_o^{under}\right)\right] \tag{29}$$

This leads to the following equation:

$$t_0^{under} = \frac{1}{\beta} \ln\left(\frac{T}{T - E_0}\right) = \frac{1}{\beta} \ln\left(\frac{2T}{2T - \hbar\omega_g}\right) \tag{30}$$

and gives rise to equation (28) for the under-damped case in which $\beta$ is small.

In the over-damped case, it is the width in deformation, which mostly determines the time-dependent behaviour of the system due to diffusion, while the velocity profile adapts very fast due to the strong coupling between intrinsic and collective degrees of freedom. Therefore, we determine the time delay $t_0$ by requiring the width in deformation of the solution of the FPE at $t = 0$ to be equal to the width of the zero-point motion $\sigma_{x0}^2 = \frac{E_0}{\omega_1^2 \cdot \mu} = \frac{\hbar}{2\mu\omega_g}$. The full analytical solution obtained in [32] for the width of the probability distribution for a parabolic potential was already given by equation (16) and approximated in the over-damped region by equation (18). As a consequence of equation (18), the width in deformation related to the zero-point motion $\sigma_{x0}^2$ is given by

$$\sigma_{x0}^2 \approx \frac{T}{\mu\omega_g^2}\left\{1 - \exp\left(-\frac{2\omega_g^2}{\beta} \cdot t_0^{over}\right)\right\} \tag{31}$$

where $t_0^{over}$ stands for the time needed to establish the zero-point motion distribution in deformation starting the calculation from a $\delta$ function.

In praxis, the temperature of the system is usually much higher than the zero-point energy (let us remind that the effective temperature which corresponds to the zero point motion amounts to 0.5 MeV in our case for which $\hbar\omega_g = 1$ MeV). This means that the width in deformation corresponding to the zero-point motion $\sigma_{x0}^2$ is much smaller than the width in thermal equilibrium $\sigma_x^2(t \to \infty)$:

$$\sigma_{x0}^2 = \frac{\hbar}{2\mu\omega_g} \ll \sigma_x^2(t \to \infty) = \frac{T}{\mu\omega_g^2} \tag{32}$$

which means that $1 - \exp\left(-\frac{2\omega_g^2}{\beta} \cdot t_0^{over}\right)$ can be replaced by $\frac{2\omega_g^2}{\beta} \cdot t_0^{over}$ in equation (31), and $t_0^{over}$ is given by

$$t_0^{over} \approx \frac{\hbar \beta}{4\omega_g T} \qquad (33)$$

It is legitimate to consider the solution of the parabolic potential in the derivation of $t_0^{over}$, since we have shown in Figure 4 that the variation of the amplitude with time for the realistic and the parabolic potential is very similar at small times.

In Figure 6, we compare the standard deviation of the probability distribution for $x < x_b$ as given by a Langevin calculation with the standard deviation $\sigma_x$ of the approximate solution obtained from equation (27). In the strongly under-damped case (Figure 6a) as well as in the strongly over-damped case (Figure 6d), the initial width of the parabolic solution including the time shift (long dashed line) agrees well with the numerical result of the Langevin calculation (thick full line). But also in the intermediate range (Figure 6b and Figure 6c), the time-shifted parabolic solution of equation (27) and the Langevin calculations come very close around $t = 0.5 \cdot 10^{-21}$ s. The deviations at smaller times are not important, since the amplitude of the probability distribution at the barrier is still so low that the flux is practically zero (see Figure 7 and the discussion below).

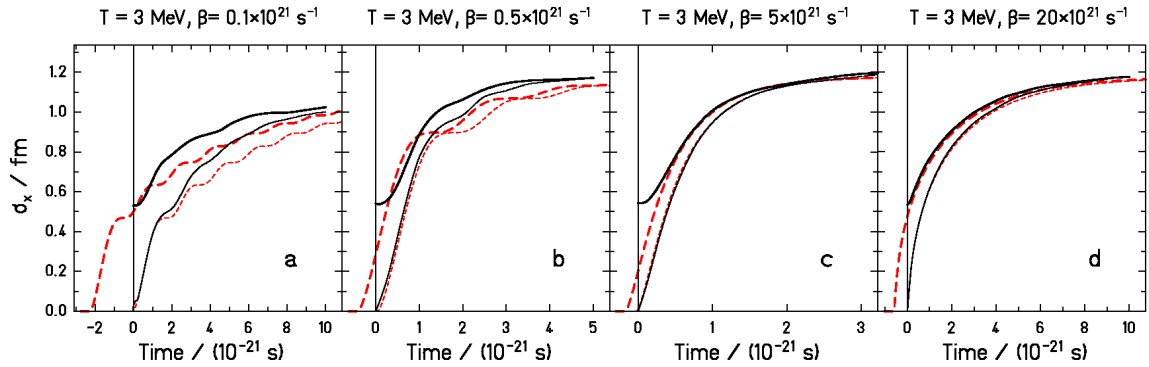

**Figure 6**: Standard deviation $\sigma_x$ of the probability distribution from a Langevin calculation inside the fission barrier of the potential given in Figure 2 (full lines) compared to the standard deviation of the analytical solution of the FPE for the parabolic potential with (long-dashed line) and without (short-dashed line) effective time shift $t_0$ given by equation (28). In all cases we have considered the nucleus $^{248}$Cm and $T = 3$ MeV. Two curves are given for the Langevin calculation: The thin full curve starts from $x = 0$ and $p = 0$. The thick full curve has been obtained by starting the trajectories with a sample from Gaussian distributions in deformation and momentum, corresponding to the zero-point motion in the parabolic potential adapted to the curvature of the potential of Figure 2 at the ground state. The comparison is given for four examples, from a strongly under-damped ($\beta = 0.1 \cdot 10^{21}$ s$^{-1}$) to a strongly over-damped case ($\beta = 20 \cdot 10^{21}$ s$^{-1}$).

The thin full lines in Figure 6 have been obtained assuming a $\delta$ function located at $x = 0$ and $p = 0$ as initial condition. The comparison between the thin and the thick full curves shows that the inclusion of the zero-point motion as initial condition in the Langevin equation corresponds essentially to a time shift. Furthermore, the analytical approximation shifted by the time lapse $t_0$ of equation (28) enables us to reproduce the width in deformation of the probability distribution corresponding to the initial conditions of the Langevin calculation, as shown by the long-dashed lines.

The influence of the introduction of $t_0$ on the fission-decay rate $\lambda_f(t)$ is illustrated in Figure 7. There it is clearly seen that shifting the analytical expression (long dashed line) by the time lapse $t_0$ nicely reproduces the onset of fission obtained directly from the solution of the equation of motion (histogram) started with the probability distribution corresponding to the zero-point motion.

To finish this section, we would like to mention that elaborate studies [35] have shown that the variance $\sigma_{x0}^2$ of the position distribution of the zero-point motion of a damped oscillator is smaller than the value $\hbar/(2\mu\omega_g)$ found for an undamped oscillator, which we have used to derive equation (33). In particular it was shown, that this difference becomes more important in the over-damped regime with increasing damping [35, 36]. Quantitatively, for $\beta=1.52\cdot10^{21}$ s$^{-1}$ the variance of the damped quantum oscillator at the ground state is 24% smaller than the variance of the undamped quantum oscillator, and for $\beta=15.2\cdot10^{21}$ s$^{-1}$ it is around 67% smaller. Since $t_0^{over}$ is proportional to $\sigma_{x0}^2$, the same reduction as for the variance applies for $t_0^{over}$. However, Figure 7 shows that the transient time is about one order of magnitude larger than the effective time shift $t_0^{over}$ and therefore the smaller variance of the damped quantum oscillator has only little effect on the time dependence of the fission-decay rate.

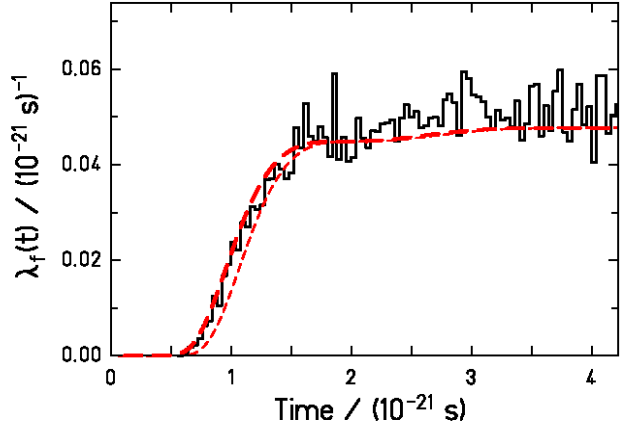

**Figure 7**: Time-dependent fission-decay rate $\lambda_f(t)$ resulting from the Langevin calculation (histogram) compared to the analytical approximation of [30] including (long dashed line, see equation (27)) or not (short dashed line, see equation (17)) the time shift $t_0$ defined by equation (28) for the nucleus $^{248}$Cm with $\beta=2\cdot10^{21}$ s$^{-1}$ at $T=3$ MeV. Note that shifting the analytical approximation in time permits to describe the onset of the process as done by the numerical solution of the equation of motion.

## 2.5. Improved approximation for the time-dependent fission width

The analytical approximation of the time-dependent fission decay width $\Gamma_f(t)$ discussed in section **2.3.** was derived neglecting the variation of the mean velocity at the barrier $\bar{v}(x=x_b;t)$. In this section we propose to include this variation in an approximate way. It can be shown that in the FPE the mean velocity and the logarithmic slope of the probability distribution at a given deformation $x$ are closely related. To demonstrate this, we will make use of the Smoluchowki equation [31]:

$$\frac{\partial}{\partial t}W(x;t) = \frac{1}{\beta\mu}\left\{\frac{\partial}{\partial x}\left[\frac{dV}{dx}W(x;t)\right] + T\frac{\partial^2}{\partial x^2}W(x;t)\right\} \quad (34)$$

where: $V$ stands for the nuclear potential (see the example in Figure 2).

After integration, regarding that $\frac{dV}{dx}=0$ at the barrier $x_b$ and that $W(x;t) = 0$ for $x \to -\infty$ as well as $\frac{\partial}{\partial x}W(x;t)=0$ for $x \to -\infty$, we obtain the following result:

$$\int_{-\infty}^{x_b}\frac{\partial}{\partial t}W(x;t)dx = \frac{T}{\beta\mu}\frac{\partial W(x;t)}{\partial x}\bigg|_{x_b} \quad (35)$$

Applying the continuity equation with the definition of the mean velocity introduced in equation (10) results in:

$$\int_{-\infty}^{x_b}\frac{\partial}{\partial t}W(x;t)dx = -\int_{-\infty}^{+\infty}v\cdot W(x=x_b,v;t)dv = -\bar{v}\big|_{x_b}\cdot\int_{-\infty}^{+\infty}W(x=x_b,v;t)dv = -\bar{v}\big|_{x_b}\cdot W(x=x_b;t)$$

(36)

Note that by the continuity equation the velocity enters explicitly into this equation, while it is only an implicit variable in the Smoluchowski equation.

By dividing the previous expression by $W(x=x_b;t)$ and using equation (35) we finally obtain:

$$\bar{v}\big|_{x_b} = -\frac{\frac{T}{\beta\mu}\frac{\partial W(x;t)}{\partial x}\big|_{x_b}}{W(x=x_b;t)} = -\frac{T}{\beta\mu}\frac{\partial}{\partial x}[ln(W(x;t))]\big|_{x_b} \quad (37)$$

As can be seen from equation (37), the mean velocity $\bar{v}\big|_{x_b} = \bar{v}(x=x_b;t)$ at the fission barrier is proportional to the logarithmic derivative of the amplitude at the fission barrier $\partial \ln(W)/\partial x\big|_{x=x_b}$. Therefore, the approximated analytical description of the fission-decay width $\Gamma_f(t)$ we proposed in ref. [30] and derived in detail in section **2.2** can be improved by introducing the variation of the logarithmic slope of the probability distribution at the barrier.

Unfortunately, we have no access to an analytical solution of the probability distribution above the realistic potential. Therefore, we use again the analytical solution of the FPE for the parabolic potential to account for the variation of the velocity profile in an approximate way. Figure 8 demonstrates with two examples that the variation in time of the logarithmic slope of the normalised amplitude at the barrier deformation is qualitatively similar for the parabolic and the realistic potential, although it quantitatively fails to reproduce the full variation. This similarity is understandable: Although the parabolic potential at $x = x_b$ is not flat, the expression for the mean velocity for the parabolic potential corresponding to equation (37) still contains a term which is proportional to the logarithmic slope of the potential. This means that one obtains an improved estimation of the time-dependent fission width when equation (26) is replaced by the following expression:

$$\Gamma_f \approx \frac{W^{par}(x=x_b;t)}{W^{par}(x=x_b;t \to \infty)} \cdot \left[ \frac{-d\ln\{W^{par}(x;t)\}}{dx} \bigg/ \frac{-d\ln\{W^{par}(x;t \to \infty)\}}{dx} \right]_{x=x_b} \cdot \Gamma_f^K \qquad (38)$$

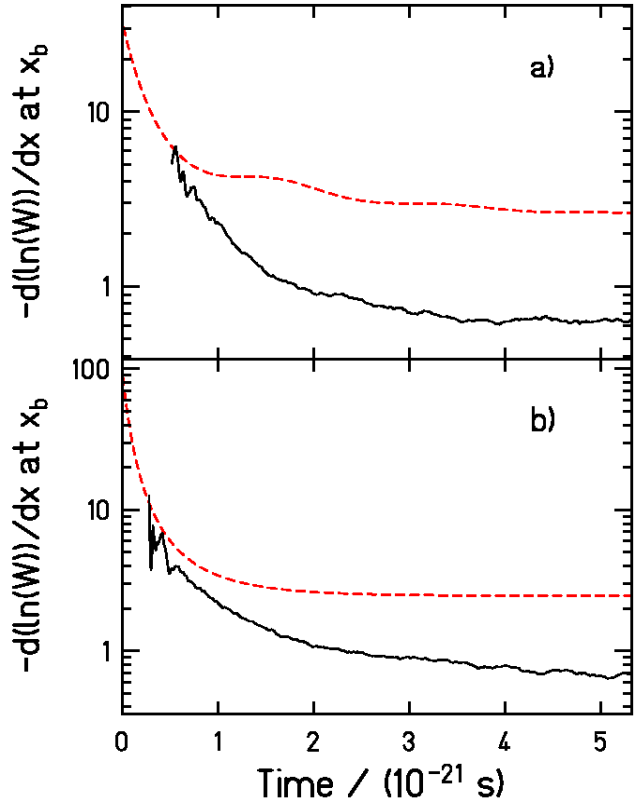

**Figure 8**: The negative logarithmic slope of the probability distribution at the barrier position $x_b$ for two cases: $T = 3\,\text{MeV}$, $\beta = 0.5 \cdot 10^{21}\,s^{-1}$ (upper part) and $T = 3\,\text{MeV}$, $\beta = 5 \cdot 10^{21}\,s^{-1}$ (lower part). The full line shows the numerical result of the Langevin approach with a realistic potential, while the dashed line corresponds to the result obtained with the analytical approximation for the parabolic potential given by equation (27).

Another deficiency of the analytical approximation proposed in ref. [30] results from the normalisation to the Kramers stationary value $\Gamma_f^K$. Indeed it is known that Kramers prediction underestimates the stationary fission width for temperatures larger than the fission barrier, leading to a discrepancy between our approximation and the exact FP or Langevin result in the stationary regime for $T \geq B_f$ (see Figure 1). To remove this deficiency, we propose to make use of the concept of the Mean First Passage Time (MFPT) [37]. The MFPT is the mean value of the distribution of first passage times. Although it can be evaluated at any deformation point $x_e$, it is physically meaningful at scission only, where the negative current of trajectories is negligible. This restriction is due to the absorption at $x_e$ in the definition of the MFPT itself. Consequently, at scission the MFPT is equivalent to the mean scission time $\tau_{scission}$. Furthermore, with the help of numerical FP calculations, Grangé et al. showed that $\tau_{scission}$ can be expressed as the sum of three contributions: the initial delay due to transient effects, the statistical decay time $\tau_{stat}$ and the dynamical saddle-to-scission time $\tau_{ss}$. Thus, they estimated the mean scission time by the following sum:

$$\tau_{scission} = \tau_{stat} + \tfrac{1}{2}\tau_{trans} + \tau_{ss} \tag{39}$$

where the delay due to transient effects is approximated by half the transient time $\tau_{trans}$ as given by equation (1), and the saddle-to-scission time $\tau_{ss}$ is taken from [26]

$$\tau_{ss} = \frac{1}{2\cdot\sqrt{\omega_{sad}^2 + \frac{\beta^2}{4}} - \beta} \cdot \ln\left|\frac{\left(2\cdot x_{sci}\cdot\sqrt{\omega_{sad}^2 + \frac{\beta^2}{4}}\right)^2 \cdot \left(\frac{\beta}{2} - \sqrt{\omega_{sad}^2 + \frac{\beta^2}{4}}\right)}{\frac{\beta T}{\mu}}\right| \tag{40}$$

The determination of the MFPT requires a numerical FP or Langevin calculation. However, in the over-damped regime, it can be evaluated analytically using the closed expression [37]:

$$MFPT = \frac{\beta\mu}{T}\int_0^{x_e} du\, \exp\left[\frac{V(u)}{T}\right]\int_{-\infty}^{u} dv\, \exp\left[-\frac{V(v)}{T}\right] \tag{41}$$

Thus, the validity of the analytical approximation of $\Gamma_f(t)$ given by equation (38) can be extended to temperatures larger than the fission barrier, for which the stationary value of Kramers cannot be used, replacing the Kramers width $\Gamma_f^K$ by the statistical fission-decay width $\Gamma_{stat}$, obtained from the following equation:

$$\Gamma_{stat} = \frac{\hbar}{\tau_{stat}} \approx \frac{\hbar}{MFPT - \tfrac{1}{2}\tau_{trans} - \tau_{ss}} \tag{42}$$

Unfortunately, the closed expression of the MFPT as expressed with equation (41) being restricted to the over-damped region, such an analytical determination of the normalisation factor $\Gamma_{stat}$ is not valid in the under-damped region. However, since the validity of the Kramers solution for the stationary fission rate is limited by the ratio $T/B_f < 1$, one could imagine that the relative deviation

$$r = \frac{\Gamma_{stat}}{\Gamma_f^K} \qquad (43)$$

also quantitatively scales with the value of $T/B_f$, independently of the value of $\beta$. Thus, we investigated the behaviour of $r$ as a function of the dissipation strength $\beta$ by dividing the statistical fission-decay width $\Gamma_{stat}$ obtained from the numerical Langevin calculation by Kramers' prediction $\Gamma_f^K$. As can be seen in Figure 9, the value of $r$ was found to vary little as a function of $\beta$ for a given temperature.

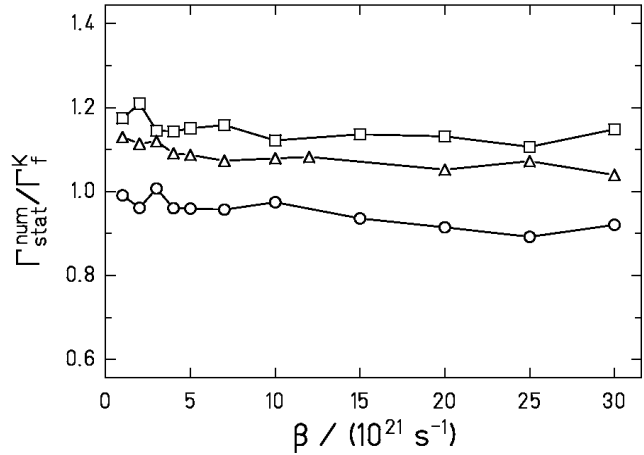

**Figure 9:** Ratio $r$ between the statistical fission-decay width $\Gamma_{stat}^{num}$ as extracted from a numerical Langevin calculation and Kramers' prediction $\Gamma_f^K$ as a function of the dissipation strength $\beta$. The calculations were performed for a $^{248}$Cm fissioning nucleus and a fission barrier of 3.7 MeV. The results are shown for three different temperatures, $T = 2$ MeV (circles), $T = 4$ MeV (triangles) and $T = 5$ MeV (squares).

Thus, we propose the following prescription to evaluate the stationary fission-decay width: With the help of equations (42) and (43), the ratio $r$ can be determined analytically in the over-damped regime, where $\Gamma_{stat} \approx \dfrac{\hbar}{MFPT - \frac{1}{2}\tau_{trans} - \tau_{ss}}$. The statistical fission-decay width in the under-damped regime can then be calculated according to equation (43) as the product of the ratio $r$ obtained for the over-damped regime and the corresponding $\Gamma_v^K$ for the under-damped regime. This solution allows overcoming the deficiency of Kramers prediction for large temperatures in the cases of over-damped *and* under-damped systems.

## 2.6. Quantitative results for the fission-decay rate

In the previous sections we have shown that the analytical approximation of the time-dependent fission-decay width proposed in ref. [30] and improved in this work, is based on realistic assumptions. In this section, we systematically investigate its qualities and shortcomings over a wide range of dissipation strength and temperature. Figure 10 compares the fission-decay rate obtained by a Langevin code (histogram), the analytical approximation of [30] (short dashed line) and the improved expression of it given in the previous section (long dashed line). The single FP curve (full line) drawn in Figure 10 for $\beta = 5 \cdot 10^{21} s^{-1}$ and $T = 5$ MeV stands only for reminding the equivalence of both Langevin and Fokker-Planck methods.

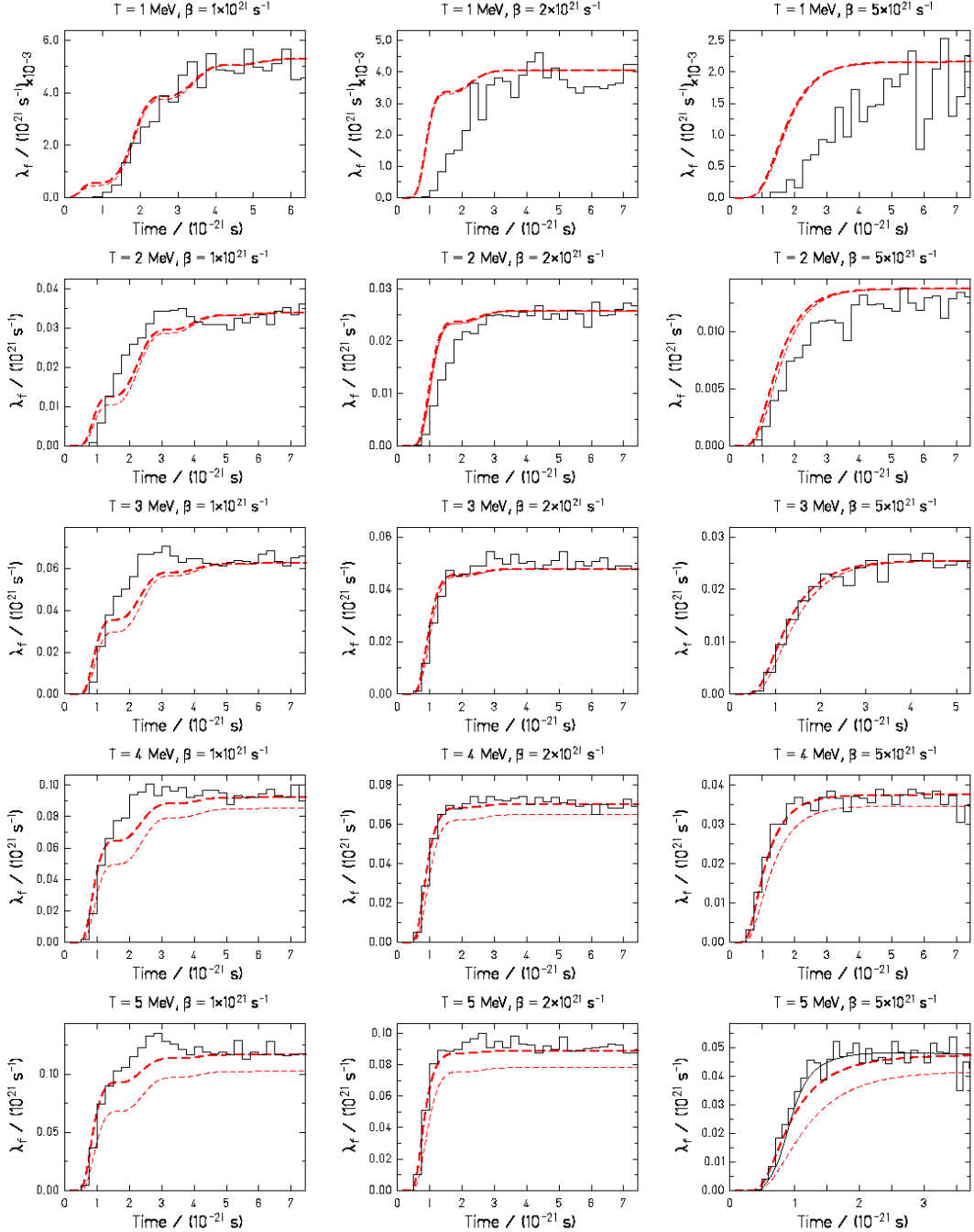

**Figure 10**: Systematic comparison of different approaches to the time-dependent fission-decay rate for different values of temperature $T$ and reduced friction coefficient $\beta$: Langevin approach (histogram), analytical approximation of [30] (short dashed line), improved expression given in this work (long dashed line). For T = 5 MeV and $\beta = 5 \cdot 10^{21}$ s$^{-1}$, the solution of the FPE (full line) is shown in addition. For the numerical Langevin and FP calculations, the deformation-dependent potential shown in Figure 2 has been used. The zero-point motion is considered by an initial effective temperature of 0.5 MeV.

Whereas the description of Bhatt et al. [26] starts too early as seen in Figure 1, the approximation of ref. [30] describes quite well the initial inhibition of fission. Under the condition that the system is not strongly under-damped and that the temperature is not too low, the analytical approximation reproduces the slope of the escape rate quite well. The agreement is less good in the under-damped region. However, as will be seen from the comparison with the experimental data, the range $1 \cdot 10^{21} s^{-1} \leq \beta \leq 5 \cdot 10^{21} s^{-1}$ is the most important. Let us note that the improved expression proposed by equation (38) leads to a slightly better agreement with the Langevin result at earliest times, and the modifications introduced by equations (41) to (43) permit to overcome the limitations of Kramers' description at temperatures larger than the fission barrier.

In ref. [26] the analytical approximation was derived separately for the under-damped and the over-damped region. The authors used the Smoluchowski equation (34) as soon as $\frac{\beta}{2\omega_g} > 1$ that defines the limit between the under-damped and over-damped regime. This equation, in which the influence of inertia is neglected, corresponds to the reduced FPE in the asymptotic case of strongly over-damped systems ($\beta \to \infty$). In ref. [26], $\omega_g = 1.83 \cdot 10^{21}$ s$^{-1}$ is used so that $\beta = 5 \cdot 10^{21}$ s$^{-1}$ belongs to the over-damped region. We compare in Figure 11 the escape rate obtained by the Langevin and Smoluchowski equations for $\omega_g = 1.83 \cdot 10^{21}$ s$^{-1}$ and for three values of $\beta$ in the over-damped regime. There it can be seen that the sufficiently over-damped regime in which the Smoluchowski equation (34) is valid starts for values of $\beta$ larger than about $10 \cdot 10^{21}$ s$^{-1}$. This observation explains the discrepancy between the Langevin calculation and the approximation of Bhatt et al. for $\beta = 5 \cdot 10^{21}$ s$^{-1}$, illustrated in Figure 1. This proves that $\beta = 5 \cdot 10^{21}$ s$^{-1}$ does not correspond to a strong enough over-damped system for the Smoluchowski equation to be valid, at least for the present purpose to study transient effects where the onset of the fission-decay width is crucial. This implies that the Smoluchowski equation is not valid in the regime of one-body dissipation which typically corresponds to $\beta \approx 5 \cdot 10^{21}$ s$^{-1}$. Although the approximation of Bhatt et al. in [26] is rather well suited in the under-damped regime (see Figure 2 of ref. [26]), its deficiency to describe the earliest stages of the process for $\beta \approx 5 \cdot 10^{21}$ s$^{-1}$ was the reason that motivated us to derive another analytical approximation to the exact solution of the FPE. Our approximation gives a uniform continuous formulation for both the under-damped and the over-damped region and reproduces the onset of fission resulting from the numerical calculations rather closely.

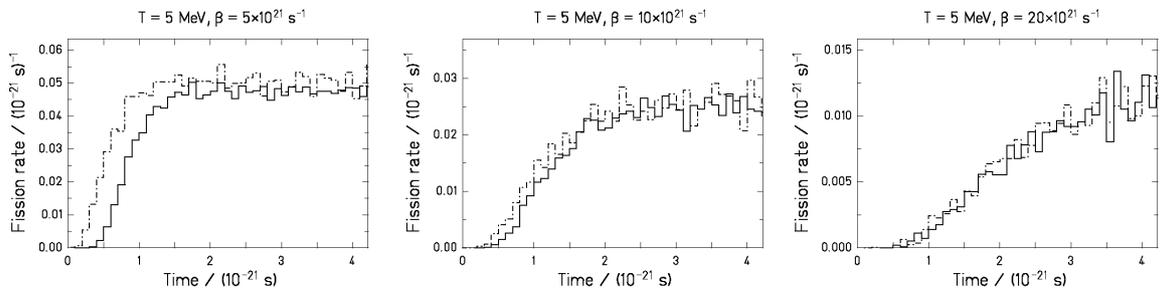

**Figure 11**: Time-dependent fission-decay rate as obtained from the resolution of the Smoluchowski equation (dash-dotted histograms) compared to the result of the Langevin equation (full histograms)

calculated with different values of $\beta$ for the nucleus $^{248}$Cm at $T$ = 5MeV and a fission barrier of 3.7 MeV.

## 3. Dynamical approaches to the nuclear de-excitation process

In general, there exist three methods to model the decay of a heavy excited nucleus in a dynamical way. One method is based on the Langevin equation [15, 16], another one on the FP equation [14]. In both cases, the evolution of the system is followed in small time steps, either by computing the individual trajectories or the integral probability distribution of the system, respectively. At each time step, the probability for the evaporation of particles is computed and randomly decided. The third option corresponds to a dynamical evaporation code, in which the fission decay width is obtained by the solution of the Langevin or FP equation *at each step* of the evaporation chain. Such a code is equivalent to the Langevin or FP treatment. Unfortunately, as already stressed in section **1**, such a procedure is inconceivable in many applications due to the high computational time required. However, any analytical approximation of the time-dependent fission decay width can replace the numerical solution, without destroying the equivalence of the code with a Langevin or FP approach, under the condition that the approximation used is as close as possible to the numerical solution. This crucial condition is fulfilled by the analytical formulation we propose in the present work, as we have shown.

In view of the above-mentioned equivalence, a statistical evaporation-fission code can be transformed into a dynamical de-excitation code by introducing a time-dependent fission-decay width. When two slightly simplifying assumptions are applied, it is enough to consider the time-dependence of the fission width. In this case, the evolution of the system in deformation does not enter explicitly in the code. Firstly, the deformation dependence of the particle-decay widths may be neglected. Secondly, the variation of the available intrinsic excitation energy as a function of deformation, investigated in reference [38], may be disregarded. Both approximations are not crucial in calculations which are restricted to the small deformation range from the initial state up to the saddle point. These effects could be considered by replacing the respective constant values by the values obtained by averaging over the actual deformation distribution in the corresponding time steps. Details on the implementation of the analytical approximation to the solution of the FPE developed in this work in our de-excitation code ABRABLA are given in reference [25].

## 4. Summary

The relaxation process of a nuclear system towards equilibrium leads to a *delay* of fission compared to the Kramers fission decay time. This feature of nuclear fission, which originates from dissipation, is automatically brought to light by solving the equation of motion in the Fokker-Planck or Langevin approach. An equivalent procedure, which in addition enables one to avoid high computational times, consists of including a realistic analytical time-dependent fission-decay width in an evaporation code. A meticulous investigation of the evolution of the probability distribution of the system in phase space all along its dynamical path permitted us to extract the main features of the relaxation process. Making use of these results, we have developed an easily calculable approximation of the time-dependent fission-decay width that is based on realistic physical assumptions. Compared to other approximations widely used in the past, our new analytical formulation has proven to reproduce rather closely the trend of the exact solution in the under- as well as in the over-damped regime. At this stage, it is more than desirable to carefully study, how the description of transient effects influences the conclusions drawn on dissipation. Indeed, such an investigation may be crucial with respect to the reliability of previous works using less realistic formulations for the time-dependence of the fission decay width, and namely the exponential in-growth function. The new analytical expression, which we propose in the present work, definitely represents an improvement in that direction.

**Acknowledgement**

We acknowledge valuable discussions with Hans Feldmeier, Anatoly V. Ignatyuk, and David Boilley. This work has been supported by the European Union in the frame of the HINDAS project under contract FIKW-CT-2000-0031 and by the Spanish MCyT under contract FPA2002-04181-C04-01. One of us (C. S.) is thankful for the financing of a one-year stay at GSI by a Humboldt fellowship. The work profited from a collaboration meeting on "Fission at finite thermal excitations" in April 2002, sponsored by the ECT* ("STATE" contract).

# Appendices

## A1: Choice of the coordinate system

Describing any physical process needs to have recourse to some coordinate system. This is particularly important for the studies of fission dynamics, which deal with the evolution of a nucleus in deformation space. In this appendix we will point out that the process can be described using a constant, coordinate-independent, mass parameter, respectively mass tensor in the case of more than one dimension, without any restriction on the physics of the problem. Also the use of a constant friction strength turns out to be close to the theoretical expectations as will be shown below.

In our work, the Langevin as well as the Fokker-Planck equation were solved by assuming that neither nuclear inertia, nor nuclear friction depends on deformation. It is the aim of this appendix to estimate how crude this assumption is. As an example, we will consider the Langevin equation.

### A1.1 Equation of motion

The dynamical evolution of a fissioning nucleus can be described by the following Langevin equation of motion for a given deformation coordinate $q$ and its conjugate momentum $p$: (for simplification we restrict ourselves to the one-dimensional Langevin equation, but it can easily be generalized to $n$ dimensions)

$$\frac{dq}{dt} = \frac{p}{\mu(q)}$$

$$\frac{dp}{dt} = \frac{1}{2}\left(\frac{p}{\mu(q)}\right)^2 \frac{d\mu(q)}{dq} - \frac{dV(q)}{dq} - \frac{\gamma(q)}{\mu(q)}p + \sqrt{D(q)}f_L(t)$$

(A1.1)

where: $\mu(q)$ and $\gamma(q)$ correspond to the nuclear inertia and friction coefficient, respectively. The driving force $-\frac{dV(q)}{dq}$ is derived from the nuclear potential $V(q)$. The Langevin random force, last term of the right-hand side of the second part of equation (A1.1), describes the fluctuating, or *Brownian*, part of the surrounding medium on the motion of the particle. In the framework of the fluctuation-dissipation theorem, the diffusion coefficient $D(q)$ can be related to friction via the Einstein relation: $D(q) = \gamma(q)T$.

Dividing equation (A1.1) by $\mu(q)$ so that to make appear the velocity $\frac{dq}{dt} = \frac{p}{\mu(q)}$, it follows:

$$\frac{d^2q}{dt^2} = -\frac{1}{\mu(q)}\frac{d\mu(q)}{dq}\frac{d^2q}{dt^2} - \frac{1}{\mu(q)}\frac{dV(q)}{dq} - \frac{\gamma(q)}{\mu(q)}\frac{dq}{dt} + \frac{1}{\mu(q)}\sqrt{D(q)}f_L(t) \qquad (A1.2)$$

Numerical calculations show that the first term on the right-hand side of equations (A1.1) and (A1.2) can be neglected. Consequently, this term will be omitted in the following.

For the simple case of an oscillator characterised by its frequency $\omega$, equation (A1.2) turns into:

$$\frac{d^2q}{dt^2} = -\omega^2 q - \frac{\gamma(q)}{\mu(q)}\frac{dq}{dt}(t) + A(t) \qquad (A1.3)$$

where the last random term of the right-hand side of (A1.2) has been replaced by $A(t)$ (which does not depend on the velocity and is assumed to change rapidly compared to the variations of the velocity [32]). The frequency $\omega(q)$ is given by:

$$\omega(q) = \sqrt{\frac{d^2V/dq^2}{\mu(q)}} = \sqrt{\frac{C}{\mu(q)}} \qquad (A1.4)$$

Remembering that the diffusion term $A(t)$ is ultimately determined by the equilibrium fluctuations [6], it clearly appears that only the two ratios $\frac{C}{\mu}$ and $\frac{\gamma}{\mu}$ of the transport coefficients govern the average motion. Note that both quantities $\sqrt{\frac{C}{\mu}}$ and $\frac{\gamma}{\mu}$ have the dimension of inverse time. Thus, these ratios are invariant to transformations of the coordinate system in contrast to the transport coefficients $\mu(q)$, $\gamma(q)$ and $C$ themselves.

### A.1.2 Coordinate transformation

As the physics does not depend on the coordinate system, whereas nuclear inertia $\mu$ as well as nuclear friction $\gamma$ do, it may be convenient to choose a system in which one of these two transport coefficients is constant. Let us assume that starting from the coordinate system $q$, we are interested in a deformation space $x$ in which the inertia coefficient is constant. The total energy $E_{tot} = E_{kin} + E_{pot}$ has to be conserved when switching from one system to the other what requires:

$$\frac{1}{2}\mu(q)\left(\frac{dq}{dt}\right)^2 + V(q) = \frac{1}{2}\mu(x)\left(\frac{dx}{dt}\right)^2 + V(x)$$

$$= \frac{1}{2}\mu(q)\left(\frac{dq}{dx}\right)^2\left(\frac{dx}{dt}\right)^2 + V(q[x])$$

(A1.5)

As the potential energy $V$ at a given deformation remains the same in both coordinate systems, it follows:

$$\mu(x) = \mu(q)\left(\frac{dq}{dx}\right)^2 \tag{A1.6}$$

Requiring that the inertia coefficient is constant and equal to $\mu_0$ in the deformation space $x$, one obtains from equation (A1.6):

$$dx = dq\sqrt{\frac{\mu(q)}{\mu_0}} \tag{A1.7}$$

Starting from a given coordinate system $q$, the numerical solution of equation (A1.7) enables us to construct another coordinate system $x$ in which $\mu$ is deformation independent. After discretisation, equation (A1.7) transforms to:

$$\Delta x = \Delta q\sqrt{\frac{\mu(q)}{\mu_0}} \tag{A1.8}$$

The correspondence between two initial coordinate values $x_0$ and $q_0$ as the starting point of the discretisation procedure may be arbitrarily chosen.

As the ratio $\beta = \frac{\gamma}{\mu}$ which describes the *damping* of the system is a physical property of the process which is invariant against a coordinate transformation, the deformation dependence of $\gamma$ in the coordinate system $x$ is given by:

$$\gamma(x) = \mu_0 \frac{\gamma(q)}{\mu(q)} \tag{A1.9}$$

In the local harmonic approximation, which has proven to be quite well suited to describe nuclear collective motion [6], the invariance of the frequency $\omega$ mentioned above relates the second derivative of the potential in the respective coordinate system to the corresponding mass parameter:

$$\omega(x) = \omega(q) \quad \Rightarrow \quad \sqrt{\frac{d^2V(x)/dx^2}{\mu(x)}} = \sqrt{\frac{d^2V(q)/dq^2}{\mu(q)}} \tag{A1.10}$$

In an analogous way, one may construct a coordinate system, in which the friction coefficient $\gamma$ becomes a constant, however, it is in general not possible to obtain both a constant mass parameter and a constant friction coefficient at the same time by any coordinate transformation.

**A1.3 Coordinate system introduced by Grangé et al.**

In our work, the Langevin calculations were performed using the cubic potential shape proposed by Grangé et al. in ref. [13] displayed in Figure 2 and assuming that the transport coefficients do not vary with deformation. The nuclear inertia is taken equal to the reduced mass $\frac{A}{4}$, and the dissipation strength $\beta = \frac{\gamma}{\mu}$ is an adjustable constant input parameter. In the present section, we check with the help of independent calculations that the deformation-parameterised potential $V(x)$ proposed in [13] is not unrealistic and, moreover, that it is quite well adapted to a coordinate system in which the inertia parameter is constant.

Lets us consider the collective deformation parameter $q$ introduced in ref. [39] based on the nuclear-shape parameterisation of Trentalange et al. [40]. In ref. [41] Pomorski et al. studied the dynamical evolution of a fissioning nucleus on the Liquid Drop Model (LDM) potential landscape by solving the Langevin equation of motion for the collective deformation parameter $q$. This model has proven quite successful since the agreement between predicted and measured neutron prescission multiplicities is rather good and this over a wide range of nuclear fissioning masses [17]. Furthermore, Pomorski et al. [41] take into account a deformation-dependent inertia calculated in the framework of the incompressible Werner-Wheeler fluid approach [5] as well as a deformation-dependent friction coefficient determined by the wall-and-window formula [4]. With the help of this model, we determined the mean fission path of a $^{248}Cm$ compound nucleus parameterised as a function of $q$. We chose an excitation energy of 165 MeV so that the model of [41] leads to a fission-barrier height of about 3.8 MeV that is close to the height obtained with the cubic potential of Grangé et al. [13]. This calculation permitted us to evaluate the deformation-dependent inertia $M(q)$ and friction $\gamma(q)$ along the mean symmetric fission path. On the basis of this result, we performed a coordinate transformation using the procedure described above by equations (A1.8) and (A1.9) requiring a constant mass equal to the reduced mass $\frac{A}{4}$. That enables us to define the potential $V$ as well as the friction coefficient $\gamma$ in a new coordinate system, which we call $x$. The potential $V(x)$, resulting from this procedure was found to be very similar to the potential introduced by Grangé et al. [13], shown in Figure 2. Approximating the two potential landscapes $V(q)$ and $V(x)$ by a parabola around their respective minimum and maximum, we evaluated and compared the frequencies $\omega(q)$ and $\omega(x)$ both in the ground state and at the barrier. We obtained that the frequencies in both coordinate systems differ by about 3-4% only.

This brief study allows us to conclude that the potential as parameterised by Grangé et al. [13] and which we widely used in our work is consistent with a coordinate system in which the mass parameter is constant.

### A.1.4 Deformation-dependence of the damping strength

Although the result of the previous section can justify our approximation of a constant mass, it does not have any consequence on the additional approximation we made concerning the deformation-independence of the friction parameter $\gamma$. Indeed, in our calculation neither $\beta = \frac{\gamma}{\mu}$ nor $\mu$ depends on deformation, and thus the friction coefficient $\gamma$ is deformation-independent as well. We would like to investigate, how crude this assumption is expected to be.

As we have already stressed in the discussion of equation (A1.3), what defines the physical process, is neither $\mu$ nor $\gamma$ but the ratio $\beta$ between both, which describes the damping of the system. In ref. [16], Fröbrich et al. studied the variation of the dissipation strength $\beta$ as a function of half the distance between the centres of mass of the emerging fission fragments for several fissioning nuclei and two different friction models, one based on the one-body wall-and-window formula and the other on the two-body viscosity theory. They showed that the dissipation strength $\beta$ does not vary drastically with deformation whatever friction approach one considers. Consequently, while our deformation-independent mass does not introduce any restriction, our single approximation of a constant friction is not crucial.

## A2: Details of the Langevin calculations

The Langevin calculations performed in this work are based on numerically solving equation (A1.1). This is done using the following discretised equations:

$$x_{i+1} = x_i + \frac{p_i}{\mu}\Delta t \tag{A2.1}$$

$$p_{i+1} = p_i - \frac{dV}{dx}\Delta t - \beta \cdot p_i \cdot \Delta t + \sqrt{\beta \cdot \mu \cdot T \cdot \Delta t} \cdot \Gamma \tag{A2.2}$$

The variables and parameters are defined in table A2.1. For details see e. g. the review of Fröbrich and Gontchar [16]. The reduced friction coefficient $\beta$ and the mass parameter $\mu$ were assumed not to vary with deformation. Following ref. [26], the value of the mass parameter was set to the reduced mass $\mu = \frac{A_{cn}}{4}m_0 = \frac{A_{cn}}{4} \cdot 0.01034 \text{ MeV}(10^{-21}\text{s/fm})^2$.

| Symbol | Quantity | Unit or value |
|--------|----------|---------------|
| X | Deformation | fm |
| P | Momentum | MeV $10^{-21}$ s / fm |
| $\mu$ | mass parameter | MeV·$(10^{-21}$ s / fm$)^2$ |
| V | Potential | MeV |
| T | Temperature | MeV |
| $\beta$ | reduced friction parameter | $10^{21}$ s$^{-1}$ |
| $\Delta t$ | time step | $0.01 \cdot 10^{-21}$ s |
| $\Gamma$ | Random variable | Gaussian distribution with variance $\sigma^2 = 2$ |

**Table A2.1:** Variables and parameters used in the Langevin calculations.

# References

[1]  H. Feldmeier, Rep. Prog. Phys. 50, 915 (1987).

[2]  J. R. Nix, A. J. Sierk, in *Proceedings of the International School Seminar on Heavy Ions Physics, Dubna, USSR, 1986,* edited by M. I. Zarubina, E. V. Ivashkevich (JINR, Dubna, 1987) 453; in *Proceedings of the South Adriatic Conference on Nuclear Physics: Frontiers of Heavy Ion Physics, Dubrovnik, Yugoslavia, 1987,* edited by N. Cindro, R. Caplar, W. Greiner (World Scientific, Singapore, 1990) 333.

[3]  D. Hilscher, H. Rossner, Ann. Phys (Paris) 17, 471 (1992).

[4]  J. Blocki, J. Randrup, W. J. Swiatecki, C. W. Tsang, Ann. Phys. 105(1977) 427.

[5]  K. T. R. Davies, A. J. Sierk, J. R. Nix, Phys. Rev. C 13(1976) 2385.

[6]  H. Hofmann, Phys. Rep. 284 (1997) 137.

[7]  H. Hofmann, *Proceedings of the RIKWN Symposium on 'Dynamics in Hot Nuclei'*, Tokyo (1998).

[8]  H. Hofmann, F. A. Ivanyuk, C. Rummel, S. Yamaji, Phys. Rev. C 64(2001) 054316-1.

[9]  N. Bohr, J. A. Wheeler, Phys. Rev. 56 (1939) 426.

[10] H. A. Kramers, Physika VII 4 (1940) 284.

[11] A. Gavron, J. R. Beene, R. L. Ferguson, F. E. Obensahain, F. Plasil, G. R. Young, G. A. Petit, M. Jaaskelainen, D. G. Sarantites, C. F. Maguire, Phys. Rev. Lett. 47 (1981) 1255. Erratum: Phys. Rev. Lett. 48 (1982) 835.

[12] D. Hilscher, E. Holub, U. Jahnke, H. Orf, H. Rossner, Proc. of the 3rd Adriatic Europhysics Conference on the Dynamics of Heavy-Ion Collisions, Hvar, Croatia, Yugoslavia, May 25-30 (1981) 225.

[13] P. Grangé, L. Jun-Qing, H. A. Weidenmüller, Phys. Rev. C 27 (1983) 2063.

[14] E. Strumberger, K. Dietrich, K. Pomorski, Nucl. Phys, A529, 522 (1991).

[15] Y. Abe, S. Ayik, P.-G. Reinhard, E. Suraud, Phys. Rep. 275 (1996) 49.

[16] P. Fröbrich, I. I. Gontchar, Phys. Rep. 292 (1998) 131.

[17] K. Pomorski, B. Nerlo-Pomorska, A. Suroviec, M. Kowal, J. Bartel, K. Dietrich, J. Richert, C. Schmitt, B. Benoit, E. de Goes Brennand, L. Donadille, C. Badimon, Nucl. Phys. A 679 (2000) 25.

[18] P. N. Nadtochy, G. D. Adeev, A. V. Karpov, Phys. Rev. C 65 (2002) 064615.

[19] J. Benlliure, P. Armbruster, M. Bernas, A. Boudard, T. Enqvist, R. Legrain, S. Leray, F. Rejmund, K.-H. Schmidt, C. Stephan, L. Tassan-Got, C. Volant, Nucl. Phys. A 700 (2002) 469.

[20] G. van 't hof, J. C. S. Baceler, I. Dioszegi, M. N. Harakeh, W. H. A. Hesselink, N. Kalantar-Nayestanaki, A. Kugler, H. von der Ploeg, A. J. M. Plompen, J. P. S. van Schagen, Nucl. Phys. A 638 (1998) 613-661.

[21] V. A. Rubchenya, A. V. Kuznetsov, W. H. Trzaska, D. N. Vakhtin, A. A. Alexandrov, I. D. Alkazov, J. Aeystoe, S. V. Khlebnikov, V. G. Lyapin, O. I. Osetrov, Yu. E. Penionzhkevich, Yu. V. Pyatkov, G. P. Tiourin, Phys. Rev. C 58 (1998) 1587.

[22] B. B. Back, D. J. Blumenthal, C. N. Davids, D. J. Henderson, R. Hermann, D. J. Hofman, C. L. Jiang, H. T. Penttilä, A. H. Wuosmaa, Phys. Rev. C 60 (1999) 044602.

[23] I. Diószegi, N. P. Shaw, I. Mazumdar, A. Hatzikoutelis, P. Paul, Phys. Rev. C 61 (2000) 024613.

[24] N. P. Shaw, I. Dioszegi, I. Mazumdar, A. Buda, C. R. Morton, J. Velkovska, J. R. Beene, D. W. Stracener, R. L. Varner, M. Thoennessen, P. Paul, Phys. Rev. C 61 (2000) 044612.

[25] B. Jurado, C. Schmitt, K.-H. Schmidt, J. Benlliure, A. R. Junghans, submitted („*Manifestation of transient effects in fission induced by relativistic heavy-ion collisions*')

[26] K.-H. Bhatt, P. Grangé, B. Hiller, Phys. Rev. C 33 (1986) 954.

[27] E. M. Rastopchin, S. I. Mul´gin, Yu. B. Ostapenko, V. V. Pashkevich, M. I. Svirin, G. N. Smirenkin, Yad. Fiz. 53 (1991) 1200 [Sov. J. Nucl. Phys. 53 (1991) 741].

[28] R. Butsch, D. J. Hofman, C. P. Montoya, P. Paul, M. Thoennessen, Phys. Rev. C 44 (1991) 1515.

[29] L. G. Moretto, Proc. Third IAEA Symp. on the Physics, Chemistry of Fission, Rochester, NY, 13-17 August 1973, vol 1 (IAEA, Vienna, 1974) p. 329.

[30] B. Jurado, K.-H. Schmidt, J. Benlliure, Phys. Lett. B 533 (2003) 186 (arXiv/nucl-ex/0212020).

[31] P. Fröbrich, R. Lipperheide, *Theory of Nuclear Reactions*, Oxford Studies in Nuclear Physics 18 (1996).

[32] S. Chandrasekhar, Rev. Mod. Phys. 15 (1943) 1.

[33] D. Boilley, Y. Abe, J. D. Bao, Eur. Phys. J. A 18 (2003) 627.

[34] J. D. Bowman, W. J. Swiatecki, C. E. Tsang, Report LBL-2908 (1973).

[35] Quantum Dissipative Systems, U. Weiss, World Scientific, Singapore, 1993, ISBN 981-02-0754-9.

[36] J. Ankerhold, P. Pechukas, H. Grabert, Phys. Rev. Lett. 87 (2001) 086802-1.

[37] P. Hänggi, P. Talkner, B. Morkovec, Rev. Mod. Phys. 62 (1990) 251.

[38] G. Chaudhuri, S. Pal, Eur. Phys. J. A 14 (2002) 287.

[39] J. Bartel, K. Mahboud, J. Richert, K. Pomorski, Z. Phys. A 354 (1996) 59.

[40] S. Trentalange, S. E. Koonin, A. Sierk, Phys. Rev. C 22 (1980) 1159.

[41] K.Pomorski, J. Bartel, J. Richert, K. Dietrich, Nucl. Phys. A 665 (1996) 87.